\documentstyle[12pt,prd,aps,epsfig,preprint]{revtex}
\begin{document}
\draft
\date{\today}
\tightenlines
\title{QUANTUM STATISTICAL EFFECTS ON FUSION DYNAMICS OF HEAVY-IONS}
\author{ S.\  Ayik$^{1},$ B. Yilmaz$^{2,3},$ A. Gokalp$^{3},$ O. Yilmaz$^{3},$ N. Takigawa$^{4}$ }
\address{$^{1}$Physics Department, Tennessee Technological University, Cookeville, TN 38505, USA}
\address{$^{2}$Physics Department, Ankara University, 06100 Ankara, Turkey}
\address{$^{3}$Physics Department, Middle East Technical University, 06531 Ankara, Turkey}
\address{$^{4}$Department of Physics, Tohoku University, 980-8578 Sendai, Japan }
\maketitle

\begin{abstract}
In order to describe the fusion of two very heavy nuclei at near
barrier energies, a generalized Langevin approach is proposed,
which incorporates the quantum statistical fluctuations in
accordance with the fluctuation and dissipation theorem. It is
illustrated that the quantum statistical effects introduce an
enhancement of the formation of compound nucleus, though the
quantum enhancement is somewhat less pronounced as indicated in
the previous calculations.
\end{abstract}

\thispagestyle{empty} ~~~~\\ 
\newpage
\setcounter{page}{1}

\section{INTRODUCTION}

In recent years, there has been a great deal of interest in the
synthesis of superheavy elements by means of heavy-ion fusion at
near barrier energies \cite{R1}. Due to very low production
probabilities, investigation of heavy-ion fusion reactions remains
a challenging task both experimentally and theoretically.
Theoretical investigations are mostly based on transport theory in
which heavy-ion fusion is viewed as a diffusion process
\cite{R2,R3,R4,R5}. During the approach phase, a part of the
kinetic energy of the relative motion dissipates into intrinsic
degrees of freedom, colliding ions overcome the Coulomb barrier
and a sticking configuration is formed. Subsequently, the system
evolves inward over the conditional saddle to form a spherical
compound nucleus. Heavy-ion fusion is described in terms of a few
relevant variables, which evolve according to a Langevin dynamics
as in a typical diffusion process. In most of these investigations
quantum statistical effects are ignored and a classical treatment
is employed, in which friction and diffusion properties are
related through classical fluctuation-dissipation theorem. Since
superheavy elements are stabilized by the shell correction energy,
they should be synthesized at reasonably low energies, which
corresponds to nuclear temperatures of the order of $T=0.5-1.0$
MeV. On the other hand, the curvature of the conditional saddle is
also of the order of $\hbar\Omega=1.0$ MeV. As a result, it is
expected that the quantum statistical effects play an important
role in the formation of a compound nucleus by diffusion along the
conditional potential barrier \cite{R6,R7,R8}.  In a recent work,
using the density matrix formalism, we derived a generalized
Fokker-Planck equation for the distribution function of relevant
collective variables with non-Markovian transport coefficients and
we illustrated the quantum statistical effects on the formation of
compound nucleus at low temperatures \cite{R9,R10}. Here, we
consider the same problem, but follow a different description
based on generalized Langevin approach.  By projection on the
collective space, it is possible to deduce stochastic equations of
motion for collective variables. The friction term involves memory
effect and the random force acts as a correlated noise and they
are related according to quantum fluctuation-dissipation theorem
\cite{R11,R12}. As a result, the quantum statistical fluctuations
are incorporated into the description and the approach is valid at
all temperatures. In principle, the Langevin approach presented
here is equivalent to the density matrix formalism, however it has
certain advantages as compared to the description by the
Fokker-Planck equation \cite{R13}. One thing is that, in realistic
situations, numerical simulations of Langevin equation require
much less numerical effort than the solution of the Fokker-Planck
equation. Furthermore, diffusion coefficients presented here are
modified by the friction mechanism, which was not considered in
the density matrix approach \cite{R9,R10}. In section 2, we derive
generalized Langevin equations for the relevant collective
variables. In section 3, we present an analysis of the Langevin
equation. In section 4, in order to illustrate the effect of
quantum statistical fluctuations, we present calculations to
describe the heavy-ion fusion reactions at low temperatures.
Finally, we give conclusions in section 5.

\section{GENERALIZED LANGEVIN EQUATIONS}

It is possible to derive transport equation for the entire
dynamics of the heavy-ion fusion process, starting from the
entrance channel until formation of a compound nucleus. Here, we
want to investigate the influence of quantum statistical
fluctuations on the formation probability of compound nucleus at
near barrier energies. For this purpose, we consider a part of the
dynamics, namely, evolution of the system along the conditional
saddle from sticking configuration until formation of compound
nucleus. For simplicity, we describe the evolution in terms of a
single collective variable, namely the elongation variable, $q$,
which approximately corresponds to the relative distance between
the colliding ions. We consider a model, in which evolution of the
system is described by a Hamiltonian of the form,
\begin{eqnarray}\label{e1}
  H=H_0+H_{coll}+V_{coup}
\end{eqnarray}
where $H_0$, $H_{coll}$ and $V_{coup}$ represent the Hamiltonian
of the intrinsic nucleonic degrees of freedom, the Hamiltonian of
the collective variable and the coupling interaction of the
collective variable with the intrinsic nucleonic degrees of
freedom, respectively. For simplicity, we consider a harmonic form
for the collective Hamiltonian
$H_{coll}=p^{2}/2M\pm{}M\Omega^{2}q^{2}/2$, where $M$ is the mass
parameter of the collective variable and $M\Omega^{2}$ denotes the
magnitude of the curvature parameter of the potential energy,
positive sign stands for a parabolic potential well and negative
sign for a parabolic potential barrier. Furthermore, we assume
that the coupling Hamiltonian has a linear form, $V_{coup}=qF$.
The classical equation of motion for the collective variable can
be deduced from $dp/dt=-(dH/dq)$, where $(..)$ denotes an average
over the intrinsic degrees of freedom, to give,
\begin{eqnarray}\label{e2}
  \frac{d}{dt}q(t)=\frac{1}{M}p(t) \quad \quad \text{and}
  \quad \quad
  \frac{d}{dt}p(t)\pm{}M\Omega^{2}q(t)=-Tr(F\rho)\,.
\end{eqnarray}
Here, the quantity on the right hand side denotes the force of the
intrinsic degrees of freedom on the collective motion. We consider
the case where  the coupling $F$ is a one-body operator. We then
need only the single-particle density matrix $\rho$ of the
intrinsic degrees of freedom  to calculate the force. As we
discuss below, temporal evolution of the single-particle density
matrix exhibits a stochastic behavior. As a result, the intrinsic
force has a fluctuating part on the top of its average value.
Here, we find it more convenient to calculate the fluctuating part
of force, which is determined by the fluctuating part of the
single-particle density matrix
$\delta\rho(t)=\rho(t)-\bar{\rho}(t)$, where the bar means taking
ensemble average. Assuming fluctuations are small, $\delta\rho(t)$
is determined by a linearized transport equation \cite{R14} around
the average $\bar{\rho}(t)$,
\begin{eqnarray}\label{e3}
i\hbar\frac{\partial}{\partial{t}}\delta\rho(t)-[\bar{h}(t),\delta\rho(t)]
  -[\delta q(t)F,\bar{\rho}(t)]=0
\end{eqnarray}
where $\bar{h}(t)=h+\bar{q}(t)F$, $h$ being the Hartree-Fock
Hamiltonian of the separated nuclei and $\delta
q(t)=q(t)-\bar{q}(t)$ denotes the fluctuation of the collective
variable around its average value $\bar{q}(t)$. The ensemble
average value of the density matrix is determined by,
\begin{eqnarray}\label{e33}
i\hbar\frac{\partial}{\partial{t}}\bar{\rho}(t)-
[\bar{h}(t),\bar{\rho}(t)]=0 \,.
\end{eqnarray}
For simplicity of derivation, we neglect the collision term on the
right hand side of eq.(\ref{e3}) and eq.(\ref{e33})
\cite{R15,R16,R17}, however subsequently, we incorporate the
damping width of single-particle states.  Starting from an initial
state $\delta\rho(s)$ at some time $s$, the formal solution of
eq.(\ref{e3}) can be given as,
\begin{eqnarray}\label{e4}
  \delta\rho(t)=-\frac{i}{\hbar}\int_{s}^{t}dt'\delta q(t')\left[G(t,t')FG^{\dag}(t,t'),%
  \bar{\rho}(t)\right]+G(t,s)\delta\rho(s)G^{\dag}(t,s)
\end{eqnarray}
where  the first term describes the effects of the perturbation
during the time interval $t-s$ with
$G(t,s)=\exp\left[-(i/\hbar)\int_{s}^{t}dt'\bar{h}(t')\right]\approx
\exp\left[-(i/\hbar)(t-s)\bar{h}(t)\right]$ as the mean-field
propagator, i.e. the propagator in the absence of thermal
fluctuation, and the second term represent the propagation of the
initial fluctuations $\delta\rho(s)$ of the intrinsic degrees of
freedom during the time interval from $s$ to $t$. Here, the
initial time $s$ does not represent the remote past, but rather it
is sufficiently close to the time $t$, so that the time interval
is much shorter than the relaxation time of the intrinsic degrees
of freedom $t-s\ll\tau_{rel}$. Hence, the neglect of correlations
due to collision term is justified in the description of
eq.(\ref{e4}). In this case, the effect of correlations enters
through the initial fluctuation term. Furthermore, we assume that
the collective motion is sufficiently slow so that the intrinsic
degrees of freedom is close to local equilibrium for each value of
the collective variable. In order to evaluate the matrix elements
of the fluctuating part of the density operator based on
eq.(\ref{e4}), we approximate the average density matrix in terms
of the instantaneous single-particle wave functions as
$\bar{\rho}(t)\approx\sum|\phi_{l}(t)\rangle
n_{l}\langle\phi_{l}(t)|$, where we neglect the off diagonal
elements. In this expression, the instantaneous wave functions
$\phi_{l}(t)=\phi_{l}[q(t)]$ are determined from
$(h+qF)|\phi_{l}(q)\rangle =\varepsilon_{l}(q)|\phi_{l}(q)\rangle$
for each value of the collective variable $q$, and $n_{i}=1/
\left[\exp\left[(\varepsilon_{i}-\varepsilon_{F})/T\right]+1\right]$
denotes the Fermi-Dirac occupation factor at a temperature $T$.
Employing the instantaneous representation, the matrix elements of
fluctuations can be expressed as,
\begin{eqnarray}\label{e4a}
  \delta\rho_{ij}(t)=-\frac{i}{\hbar}\int_{s}^{t}dt'\delta q(t')G_{ij}(t,t')
  \langle i|F|j\rangle (n_j - n_i)+G_{ij}(t,s)\delta\rho_{ij}(s)
\end{eqnarray}
with $G_{ij}(t,s)=\exp[-i(t-s)(\varepsilon_i -\varepsilon_j)/\hbar]$.
  It is not possible to determine detailed structure of initial
fluctuations of the intrinsic degrees of freedom. Therefore, it is
plausible to assume that each matrix element of $\delta\rho(s)$ is
a Gaussian random quantity with zero mean
$\overline{\langle{}i|\delta\rho|j\rangle{}}=0$ and a second
moment. In accordance with the fluctuation-dissipation relation of
the single-particle density matrix, we specify the second moment
of $\delta\rho(s)$ as,
\begin{eqnarray}\label{e5}
\overline{\langle{}i|\delta\rho|j\rangle{}\langle{}j'|\delta\rho|i'\rangle{}}=%
\delta_{ii'}\delta_{jj'}\frac{1}{2}\left[n_{i}(1-n_{j})+n_{j}(1-n_{i})\right]\,.
\end{eqnarray}
In the special case of diagonal elements this formula gives the
known result for fluctuations of occupation numbers,
$\overline{\langle{}i|\delta\rho|i\rangle{}\langle{}i|\delta\rho|i\rangle{}}=%
n_{i}(1-n_{i})$ \cite{R18}. Substituting eq.(\ref{e4}) into the
right hand side of eq.(\ref{e2}), we find a generalized Langevin
equation for the fluctuations of the collective variable,
\begin{eqnarray}\label{e6}
  \frac{d}{dt}\delta p(t)\pm{}M\Omega^{2}\delta q(t)=\int_{s}^{t}dt'\gamma(t-t')
\delta q(t')+\xi(t)
\end{eqnarray}
where the memory kernel in the retarded force and the random force
term are given by
\begin{eqnarray}\label{e7}
  \gamma(t-t')=\frac{i}{\hbar}\Sigma |\langle{}i|F|j\rangle|^{2}
G_{ji}(t, t')\left[n_{i}(1-n_{j})-n_{j}(1-n_{i})\right]
\end{eqnarray}
and
\begin{eqnarray}\label{e8}
  \xi(t)=-\Sigma\langle{}i|F|j\rangle%
  G_{ji}(t,s) \langle{}j|\delta\rho{}|i\rangle.
\end{eqnarray}

Using eq.(\ref{e5}), the auto-correlation function of the random
force can be expressed as,
\begin{eqnarray}\label{e9}
  \overline{\xi(t)\xi(t')}=\Sigma|\langle{}i|F|j\rangle|^{2}%
  G_{ij}(t, t') \frac{1}{2}
  \left[n_{i}(1-n_{j})+n_{j}(1-n_{i})\right]\,.
\end{eqnarray}
Dissipation and fluctuation aspects of dynamics are closely
connected to each other, the similarity of expressions for the
correlation function and the memory kernel reflects this fact. If
the decay time of the memory kernel is sufficiently short, we can
explicitly incorporate the memory effect into the retarded force
in eq.(\ref{e6}). For evolution over a short time interval from
$t'$ to $t$, by neglecting the right hand side of eq.(\ref{e6}),
we find the following relation,
\begin{eqnarray}\label{e10}
  \delta q(t')\approx{}C(t-t') \delta q(t)-S(t-t') \delta p(t)\,.
\end{eqnarray}
For a parabolic potential well, propagators $C(t-t')$ and
$S(t-t')$ are given by,
\begin{eqnarray}\label{e11}
  C(t-t')=\cos{\Omega(t-t')} \quad \quad%
  \text{and} \quad \quad
  S(t-t')=\frac{1}{M\Omega}\sin\Omega(t-t')\,.
\end{eqnarray}
On the other hand, for a parabolic potential barrier, these
propagators are given by,
\begin{eqnarray}\label{e12}
  C(t-t')=\cosh{\Omega(t-t')} \quad \quad%
  \text{and} \quad \quad
  S(t-t')=\frac{1}{M\Omega}\sinh\Omega(t-t')\,.
\end{eqnarray}
The first term in eq.(\ref{e10}) involving $\delta q(t)$
introduces a shift in the curvature parameter of the potential.
Here, we neglect this effect and substitute the second term on the
right hand side of eq.(\ref{e10}) into the right hand side of
eq.(\ref{e6}). Note that, since the fluctuations are linear, the
same equation as eq.(\ref{e6}), but without the last term on the
right hand side, holds for the average evolution by replacing
$\delta p(t)$ and $\delta q(t)$ with $\bar{p}(t)$ and
$\bar{q}(t)$, respectively. Therefore, combining the average
evolution with the fluctuations, we obtain a generalized Langevin
equation for the actual variables, $p(t)={\bar p}(t)+\delta p(t)$
and $q(t)={\bar q}(t)+\delta q(t)$,
\begin{eqnarray}\label{e13}
  \frac{d}{dt}p(t)\pm{}M\Omega^{2}q(t)=-\beta p(t)+\xi(t)\,.
\end{eqnarray}
Here, the reduced friction coefficient is given by,
\begin{eqnarray}\label{e14}
  \beta=\frac{i}{\hbar}\int_{0}^{t-s}d\tau\Sigma|\langle{}i|F|j\rangle|^{2}%
  e^{-\frac{i}{\hbar}\tau
  (\varepsilon_{j}-\varepsilon_{i})}n_{i}(1-n_{j})S(\tau)+c.c.\,.
\end{eqnarray}
Substituting eqs.(\ref{e11}) and (\ref{e12}) for $S(\tau)$, we
find for the friction coefficient for a parabolic well,
\begin{eqnarray}\label{e15}
  \beta=-\frac{1}{2iM \Omega}\Sigma |\langle i|F|j\rangle |^{2}
   && \left[\frac{e^{-\frac{i}{\hbar}(t-s)(\varepsilon_{j}-\varepsilon_{i}-\hbar\Omega-i\eta)}-1}
     {\varepsilon_{j}-\varepsilon_{i}-\hbar\Omega-i\eta}\right. \nonumber \\
   && \left.~~- \frac{e^{-\frac{i}{\hbar}(t-s)\left(\varepsilon_{j}-\varepsilon_{i}+\hbar\Omega-i\eta\right)}-1}
    {\varepsilon_{j}-\varepsilon_{i}+\hbar\Omega-i\eta}\right]n_{i}(1-n_{j})+c.c.
\end{eqnarray}
and for a parabolic barrier,
\begin{eqnarray}\label{e16}
  \beta=+\frac{1}{2M\Omega}\Sigma|\langle{}i|F|j\rangle|^{2}
  && \left[\frac{e^{-\frac{i}{\hbar}(t-s) (\varepsilon_{j}-\varepsilon_{i}-i\hbar\Omega-i\eta)}-1}
     {\varepsilon_{j}-\varepsilon_{i}-i\hbar\Omega-i\eta}\right.
     \nonumber\\
  &&~~\left. -\frac{e^{-\frac{i}{\hbar}(t-s)\left(\varepsilon_{j}-\varepsilon_{i}+i\hbar\Omega-i\eta\right)}-1}
     {\varepsilon_{j}-\varepsilon_{i}+i\hbar\Omega-i\eta}\right]n_{i}(1-n_{j})+c.c.\,.
\end{eqnarray}
In obtaining these results, we include the damping width $\eta$ of
the single-particle states into the propagator in eq.(\ref{e14})
\cite{R19}. In further evaluation of the friction coefficients, we
neglect the time-dependent terms in eqs.(\ref{e15}) and
(\ref{e16}). The reason is the following: the dominant
contributions to the friction coefficient arise from the coupling
matrix element over an energy interval of the order of major shell
spacing, $\varepsilon_{j}-\varepsilon_{i}=\Delta\approx10$ MeV,
which is much larger than typical values of collective frequency
we consider here, $\hbar\Omega\approx 1.0$ MeV. If the
single-particle spectrum is sufficiently dense, the summations
over the single particle states can be converted to energy
integrals. As a result, exponential factors in eq.(\ref{e15}) and
eq.(\ref{e16}) damp out over a time interval of the order of
$\tau_{0}=\hbar{}/\Delta$. Furthermore, in particular for low
frequency collective motion, $\hbar\Omega\leq\eta$, these
exponential factors damp out even over a shorter time scale as a
result of damping of the single-particle states. Therefore, for a
sufficiently long time interval, $t-s\gg\tau_0$, neglecting time
dependent terms, we have for a parabolic well,
\begin{eqnarray}\label{e17}
  \beta=\Sigma|\langle{}i|F|j\rangle|^{2}\frac{1}{M\Omega}
  \left[\frac{\eta}
  {\left(\varepsilon_{j}-\varepsilon_{i}-\hbar\Omega\right)^{2}+\eta^{2}}-
   \frac{\eta}
  {\left(\varepsilon_{j}-\varepsilon_{i}
  +\hbar\Omega\right)^{2}+\eta^{2}}\right]
  n_{i}(1-n_{j})
\end{eqnarray}
and for a parabolic barrier,
\begin{eqnarray}\label{e18}
  \beta=\Sigma|\langle{}i|F|j\rangle|^{2}\frac{1}{M\Omega}
  \left[\frac{\varepsilon_{j}-\varepsilon_{i}}
  {(\varepsilon_{j}-\varepsilon_{i})^{2}+(\hbar\Omega-\eta)^{2}}-
   \frac{\varepsilon_{j}-\varepsilon_{i}}
  {(\varepsilon_{j}-\varepsilon_{i})^{2}+(\hbar\Omega+\eta)^{2}}\right]
  n_{i}(1-n_{j})\,.
\end{eqnarray}
As seen from these results, for finite $\Omega$, the friction
coefficient has different expressions around a well and a barrier.
However, in the limit $\Omega\rightarrow{}0$, it can be easily
seen that these expressions become identical, known as the
one-body friction formula. We call this limiting value as the
classical friction coefficient and denote as $\beta_{0}$. We
introduce the Fourier transform of the correlation function of the
random force \cite{R20},
\begin{eqnarray}\label{e19}
  \overline{\xi(t)\xi(t')}=\int_{-\infty}^{+\infty}\frac{d\omega}{2\pi}
  e^{-i\omega (t-t')}\frac{\hbar\omega}{2T}\coth\frac{\hbar\omega}{2T}
  \cdot{}2D(\omega)
\end{eqnarray}
where,
\begin{eqnarray}\label{e20}
  D(\omega)=T\Sigma|\langle{}i|F|j\rangle|^{2}\frac{1}{\omega}
  \left[\frac{\eta}
  {\left(\varepsilon_{j}-\varepsilon_{i}
  -\hbar\omega\right)^{2}+\eta^{2}}-
   \frac{\eta}
  {\left(\varepsilon_{j}-\varepsilon_{i}
  +\hbar\omega\right)^{2}+\eta^{2}}\right]
  n_{i}(1-n_{j})\,.
\end{eqnarray}
At low frequencies, $D(\omega)$ is just
$D(\omega\rightarrow{}0)=D_{0}=MT\beta_{0}$ the classical
diffusion coefficient. On the other hand, the high frequency
behavior is restricted by the magnitude of the coupling matrix
elements. If the single particle spectrum is sufficiently dense,
the magnitude of coupling matrix elements must decrease as a
function of energy difference, mainly due to the mismatch of the
overlap of the wave functions. We can represent this behavior by a
Gaussian or a Lorentzian function
$\langle{}i|F|j\rangle^{2}\propto
\exp[-(\varepsilon_{j}-\varepsilon_{i})^{2}/2\Delta^{2}]$ or
$\propto{}1/\left[1+(\varepsilon_{j}-\varepsilon_{i})^{2}/2\Delta^{2}\right]$.
Furthermore, because of the Lorentzian factors in eq.(\ref{e20}),
we can replace the energy difference
$\varepsilon_{j}-\varepsilon_{i}$ with the frequency $\hbar\omega$
and approximately describe frequency dependence of diffusion
coefficient as
$D(\omega)=D_{0}\exp\left[-(\hbar\omega)^{2}/2\Delta^{2}\right]$,
here we take the Gaussian for the frequency spectrum. As a result,
the correlation function eq.(\ref{e19}) of the random force can be
expressed as,
\begin{eqnarray}\label{e21}
  \overline{\xi(t)\xi(t')}=2D_{0}\cdot\chi(t-t')
\end{eqnarray}
where
\begin{eqnarray}\label{e22}
  \chi(t-t')=\int_{-\infty}^{+\infty}\frac{d\omega}{2\pi}
  e^{-i\omega{}(t-t')}\frac{\hbar\omega}{2T}\coth\frac{\hbar\omega}{2T}
  \cdot\exp\left[-(\hbar\omega)^{2}/2\Delta^{2}\right]\,.
\end{eqnarray}
The correlation function is characterized by two different
parameters, the cut-off energy and temperature. Figure 1 shows the
correlation function versus time at different temperatures $T=0.5$
MeV, $1.0$ MeV and $5.0$ MeV. The results presented in this work
are not very sensitive to the cut-off energy over a range of
values $\Delta= 10 - 20 $ MeV. Therefore, in this figure and all
other figures, we employ $\Delta=15$ MeV for the cut-off energy.
At relatively high temperature $\hbar\omega\ll{}2T$,
$(\hbar\omega/2T)\coth(\hbar\omega/2T)\approx1$, and the
correlation function reduces to its classical form,
\begin{eqnarray}\label{e23}
  \chi_{0}(t-t')=\int_{-\infty}^{+\infty}\frac{d\omega}{2\pi}
  e^{-i\omega{}(t-t')}\exp\left[-\frac{(\hbar\omega)^{2}}{2\Delta^{2}}\right]
  =\frac{1}{\sqrt{2\pi}\tau_0}\exp\left[-(t-t')^{2}/2\tau_{0}^{2}\right]\,.
\end{eqnarray}
For sufficiently short decay time $\tau_0$, it can be approximated
by a delta function, $\chi_{0}(t-t')\rightarrow\delta(t-t')$, and
as a result at high temperatures, the temporal evolution becomes
Markovian and the random force $\xi(t)$ acts like a white noise.
However, as seen from the figure, in the quantal regime i.e. at
low temperatures $\hbar\Omega\geq2T$, we are faced with a
stochastic evolution with a correlated noise \cite{R11,R12}.

\section{ANALYSIS OF LANGEVIN EQUATION}

In order to obtain the joint probability distribution function
$P(q,p,t)$ of the collective variable and its conjugate momentum
$(q,p)$ by numerical simulation of the Langevin equation, in
general, we need to generate a sufficiently large ensemble of
trajectories. Since, we have a correlated noise problem, we cannot
use the standard methods \cite{R17,R21,R22} and we need to adopt
suitable algorithms for numerical simulations \cite{R23}. However,
in the situation that we consider here, the solution of the
Langevin eq.(\ref{e13}) can be given analytically \cite{R13}.
Since the equation is linear with a Gaussian random source, the
probability distribution $P(q,p,t)$ of collective variables is
also Gaussian, which is determined by the mean values
$\overline{q}(t)$, $\overline{p}(t)$ and the variances
$\sigma_{qq}(t)=\overline{\delta{}q(t)\delta{}q(t)}$,
$\sigma_{qp}(t)=\overline{\delta{}q(t)\delta{}p(t)}$,
$\sigma_{pp}(t)=\overline{\delta{}p(t)\delta{}p(t)}$ of collective
variables according to,
\begin{eqnarray}\label{e24}
  P(q,p,t)=\frac{1}{2\pi{}X}\exp\left[-\frac{1}{2X^{2}}
  \left((q-\overline{q})^{2}\tilde{\sigma}_{qq}
  +2(q-\overline{q})(p-\overline{p})\tilde{\sigma}_{qp}
  +(p-\overline{p})^{2}\tilde{\sigma}_{pp}\right)\right]
\end{eqnarray}
where $X^{2}=\sigma_{qq}\sigma_{pp}-\sigma_{qp}^{2}$ and
$\tilde{\sigma}_{ij}$ is the inverse of the $2\times2$ matrix
$(\sigma_{ij})$ with elements $\sigma_{11}=\sigma_{qq}$,
$\sigma_{12}=\sigma_{qp}$, $\sigma_{21}=\sigma_{pq}$ and
$\sigma_{22}=\sigma_{pp}$.
   The mean values of collective variables
$\overline{q}(t)$, $\overline{p}(t)$ are determined by the
classical equations of motion,
\begin{eqnarray}\label{e25}
  \frac{d}{dt}\overline{q}(t)=\frac{1}{M}\overline{p}(t) \quad
  \quad \text{and}\quad\quad \frac{d}{dt}\overline{p}(t)\pm{}
  M\Omega^{2}\overline{q}(t)=-\beta\overline{p}(t)\,.
\end{eqnarray}
Equations for variances are deduced from the Langevin equations
for the fluctuating quantities $\delta{}q(t)=q(t)-\overline{q}(t)$
and $\delta{}p(t)=p(t)-\overline{p}(t)$,
\begin{eqnarray}\label{e26}
  \frac{d}{dt}\delta{}q(t)=\frac{1}{M}\delta{}p(t) \quad
  \quad \text{and}\quad\quad \frac{d}{dt}\delta{}p(t)\pm{}
  M\Omega^{2}\delta{}q(t)=-\beta\delta{}p(t)+\xi(t)\,.
\end{eqnarray}
Multiplying both sides of these equations by $\delta{}q(t)$,
$\delta{}p(t)$ and performing ensemble averaging, we find
\begin{eqnarray}\label{e27}
  \frac{d}{dt}\sigma_{qq}(t)=\frac{2}{M}\sigma_{qp}(t)
\end{eqnarray}
\begin{eqnarray}\label{e28}
  \frac{d}{dt}\sigma_{qp}(t)\pm{}M\Omega^{2}\sigma_{qq}(t)=
  \frac{1}{M}\sigma_{pp}(t)-\beta\sigma_{qp}(t)+D_{qp}(t)
\end{eqnarray}
\begin{eqnarray}\label{e29}
  \frac{d}{dt}\sigma_{pp}(t)\pm{}2M\Omega^{2}\sigma_{qp}(t)=
  -2\beta\sigma_{pp}(t)+2D_{pp}(t)
\end{eqnarray}
where $D_{pp}(t)=\overline{\delta{}p(t)\xi(t)}$ and
$D_{qp}(t)=\overline{\delta{}q(t)\xi(t)}$ denotes the momentum and
mixed diffusion coefficients, respectively. In order to evaluate
diffusion coefficients, we need to calculate the dynamical
fluctuations of collective variables in terms of the random force.
This is carried out in Appendix A. Using the results for
$\delta{}p(t)=\int_{0}^{t}dt'Q(t-t')\xi(t')$ and from (A5) and
(A6), diffusion coefficients can be expressed in terms of the
correlation function of the random force as,
\begin{eqnarray}\label{e30}
  D_{pp}(t)=\int_{0}^{t}dt'Q(t-t')\cdot
  \overline{\xi(t')\xi(t)}=2D_{0}\int_{0}^{t}ds\,Q(s)\cdot\chi(s)
\end{eqnarray}
and
\begin{eqnarray}\label{e31}
  D_{qp}(t)=\int_{0}^{t}dt'S(t-t')\cdot
  \overline{\xi(t')\xi(t)}=2D_{0}\int_{0}^{t}ds\,S(s)\cdot\chi(s)\,.
\end{eqnarray}
In these expressions, the initial time is taken to be zero for
convenience, $t_{0}=0$, and the propagators $Q(s)$ and $S(s)$
associated with collective variables are given by (A7) and (A8) in
Appendix A. At sufficiently high temperatures, correlation
function $\chi(s)$ can be approximated by a delta function, and
consequently, the momentum diffusion coefficient is time
independent and takes its classical value, $D_{pp}=D_{0}$ and
furthermore the mixed diffusion coefficient vanishes, $D_{qp}=0$.
The mixed diffusion coefficient is a genuine non-Markovian term,
and it is absent in the Markovian limit. At low temperatures, due
to the non-Markovian behavior of the correlation function,
diffusion coefficients become time dependent and their magnitude
are strongly modified by the quantum statistical fluctuations. We,
also, note that the modified frequency
$\overline{\Omega}=\sqrt{\Omega^{2}+(\beta/2)^{2}}$ enters in
propagators $Q(s)$ and $S(s)$. For typical values of the frequency
parameter $\hbar\Omega\approx1.0$ MeV of the conditional saddle
and the magnitude of the reduced friction coefficient
$\hbar\beta/2\approx1.7$ MeV are comparable. As a result, the
friction coefficient introduces a sizable modification of
diffusion coefficients, which was not incorporated in the previous
investigation \cite{R9,R10}. Figures 2 and 3 show the diffusion
coefficients in units of $D_0$, i.e., $D_{pp}/D_{0}$ and
$D_{qp}/D_{0}$,  as a function of time for different values of
temperature, $T=0.5$ MeV, $T=1.0$ MeV and $T=5.0$ MeV. In order to
illustrate the effect of friction on diffusion coefficients, we
calculate diffusion coefficients by replacing $\overline{\Omega}$
with $\Omega$ in the propagators $Q(s)$ and $S(s)$. Figures 4 and
5 compare two different values of diffusion coefficients
$D_{pp}/D_{0}$ and $D_{qp}/D_{0}$ calculated with
$\overline{\Omega}$ and $\Omega$ as a function of time at
temperature $T=1.0$ MeV. The variances $\sigma_{qq}$,
$\sigma_{qp}$ and $\sigma_{pp}$ can be determined by solving the
coupled differential equations (\ref{e27}), (\ref{e28}) and
(\ref{e29}). However, it is much easier to obtain these variances
directly from the Langevin equation (\ref{e13}) with the help of
one-sided Fourier transform \cite{R18}, as shown in Appendix A.

\section{QUANTUM STATISTICAL EFFECTS ON DIFFUSION ALONG
CONDITIONAL SADDLE TOWARDS FUSION}

In this section, we apply the generalized Langevin approach to
investigate the influence of quantum-statistical fluctuations on
diffusion along the fusion barrier, i.e. the formation probability
$P_{f}(t)$ of compound nucleus. When the conditional saddle, i.e.
the inner fusion barrier, is approximately represented by an
inverted parabola, the formation probability, i.e. the probability
to cross the saddle point, can be calculated analytically in terms
of distribution function of the elongation parameter $q$ as
\cite{R5,R24,R25},
\begin{eqnarray}\label{e32}
  P_{f}(t)=\int_{0}^{\infty}dq\frac{1}{\sqrt{2\pi\sigma_{qq}(t)}}
  \exp\left[-\frac{(q-\overline{q}(t))^{2}}{2\sigma_{qq}(t)}\right]
  =\frac{1}{2}\text{erfc}\left(-\frac{\overline{q}(t)}
  {\sqrt{2\sigma_{qq}(t)}}\right).
\end{eqnarray}
Here, $\overline{q}(t)$ and $\sigma_{qq}(t)$ are the mean value
and the variance of the elongation parameter, which are given by
(A10) and (A11) in Appendix A. In these expressions,
$(\overline{q}_{0},\overline{p}_{0})$ are the mean values of
elongation parameter and its conjugate momentum, and
$(\sigma_{q0},\sigma_{p0})$ are the associated variances at the
initial configuration. As it is stated already, during the
approach phase of the collision, system overcome the Coulomb
barrier and  some of the initial kinetic energy is dissipated into
internal excitations and a sticking configuration is formed. The
quantities $(\overline{q}_{0},\overline{p}_{0})$ denote the
average values of the elongation parameter and its momentum at the
sticking configuration. In the second stage of the process, the
shape of the system evolves from a sticking di-nuclear
configuration towards formation of a spherical compound nucleus or
re-separate again. The asymptotic value
$P_{f}(t\rightarrow\infty)$ gives the transmission probability
from a di-nuclear configuration to compound nucleus. In order to
compare the results with our previous calculations \cite{R10}, we
consider collision of $^{48}Ca$ and $^{238}U$ nuclei and adopt the
same value for the reduced friction coefficient and the curvature
parameter of the conditional saddle to be $\beta=5\times10^{21}
s^{-1}$ and $\hbar\Omega=1.0$ MeV. We choose the initial position
$q_0$ to make the height of the conditional saddle to be 4.0 MeV
and neglect the dispersion, i.e. $(\sigma_{q0},\sigma_{p0})$, in
the initial configuration. In the classical limit, the variance
$\sigma_{qq}(t)$ of the elongation parameter has a analytical
expression given by (A14), while in the quantum limit it is given
by (A18), and involves a one dimensional numerical integration
over the frequency $\omega$. Figure 6 shows the formation
probability $P_{f}(t\rightarrow\infty)$ of compound nucleus as a
function of the initial kinetic energy
$K_{0}=\overline{p}_{0}^{2}/2M$ relative to the fusion barrier
$V_B$ at temperatures $T=0.5$ MeV, $T=1.0$ MeV and $T=5.0$ MeV.
These results, which are not very sensitive to the cut-off factor
$\Delta$, are presented for $\Delta=15$ MeV. Solid lines and
dashed lines show the quantum and the classical calculations,
respectively. At low temperatures, the quantum statistical
fluctuations give rise to an enhancement of the formation
probability, which is relevant to synthesis of superheavy elements
by heavy-ion fusion reactions. The quantum enhancement is slightly
less pronounced than that in the previous calculations \cite{R10}.
The difference  arises from the fact that in previous calculations
the mixed diffusion coefficient $D_{qp}$, which is a genuine
non-Markovian term, is neglected and the momentum diffusion
coefficient $D_{pp}$ is calculated with the unperturbed frequency
$\Omega$, rather than $\overline{\Omega}$.

\section{CONCLUSIONS}

Since several years, much effort has been made for synthesis of
superheavy elements by heavy-ion fusion at near barrier energies.
Even though, the reaction mechanism of heavy-ion fusion is not
well understood, in most theoretical descriptions fusion is viewed
as a diffusion process, which can be described by a Fokker-Planck
approach or a stochastic Langevin approach
\cite{R2,R3,R4,R5,R24,R25}. In these descriptions the quantum
statistical effects are ignored and a classical treatment is
employed. In a recent work, we introduced a description based on a
generalized Fokker-Planck approach, which incorporates quantum
effects through non-Markovian transport coefficients
\cite{R9,R10}. In the present work, we follow a different
description based on a generalized Langevin approach. The friction
term and the random force both involve memory effects and they are
related to each other in accordance with the
fluctuation-dissipation theorem. As a result, the quantum
statistical fluctuations are incorporated into the description. In
principle, both approaches provide an equivalent description,
however the Langevin approach has certain advantages in practical
applications. In general, for a complex potential energy surface,
it is much easier to carry out simulations of Langevin equation
than solving the Fokker-Planck equation. Furthermore, diffusion
coefficients presented here are strongly modified by the friction
mechanism, which was not considered in the Fokker-Planck approach
\cite{R10}. In this work, we consider a simple model in which the
fusion barrier is represented by an inverted parabola. In this
case, the joint distribution function of the collective variable
and the conjugate momentum becomes a Gaussian, and the fusion
probability can be given in an analytical form. Calculations
illustrate that the quantum statistical effects enhance the fusion
probability at low temperatures. However, the enhancement is
somewhat less pronounced than that reported in the previous
investigation \cite{R10}. For a realistic potential energy
surface, analytical solution is not possible and distribution
function of collective variables should be constructed by
generating sufficient number of events of the generalized Langevin
equation. Since, the random force is not a white noise, we cannot
use standard methods for numerical simulations \cite{R21,R22}.
Therefore, it is necessary to develop suitable algorithms for
simulations of the Langevin equation with a correlated random
force \cite{R23}.
\\
\\
\begin{center}
ACKNOWLEDGEMENTS
\end{center}
Two of us (S.A. and N. T.) gratefully acknowledges the Physics
Department of Middle East Technical University for warm
hospitality extended to them during their visits. We also thank K.
Washiyama for fruitful discussions. This work is supported in part
by the US DOE Grant No. DE-FG05-89ER40530.

\appendix
\section*{}

In this Appendix, we analyze solutions of the Langevin
eq.(\ref{e13}) together with $dq/dt=p/M$ by employing one-sided
Fourier transform \cite{R18}. After performing the Fourier
transform, we obtain
\begin{eqnarray}\label{A1}
  -q_{0}-i\omega{}q(\omega)=\frac{p(\omega)}{M}\quad\quad\text{and}
  \quad\quad -p_{0}-i\omega{}p(\omega)\pm{}M\Omega^{2}q(\omega)
  =-\beta{}p(\omega)+\xi(\omega)
\end{eqnarray}
where $(q_{0},p_{0})$ are the initial conditions,
$q(\omega)=\int_{0}^{\infty}dt\exp(i\omega{}t)q(t)$ is the
one-sided Fourier transform of the coordinate. The $p(\omega)$ and
$\xi(\omega)$ are similarly defined. Combining two equations in
(A1), the Fourier transforms of the collective variables are given
by,
\begin{eqnarray}\label{A2}
  q(\omega)=iq_{0}\frac{\omega+i\beta}{\omega^{2}\mp\Omega^{2}+i\omega\beta}
  -\frac{1}{M}\frac{p_{0}+\xi(\omega)}{\omega^{2}\mp\Omega^{2}+i\omega\beta}
\end{eqnarray}
\begin{eqnarray}\label{A3}
  p(\omega)=\frac{\pm{}M\Omega^{2}q_{0}}{\omega^{2}\mp\Omega^{2}+i\omega\beta}
  +i\omega\frac{p_{0}+\xi(\omega)}{\omega^{2}\mp\Omega^{2}+i\omega\beta}\,\cdot
\end{eqnarray}
Time dependence of the collective variables are found by the
inverse Fourier transformation,
\begin{eqnarray}\label{A4}
   q(t)=\int_{-\infty}^{+\infty}\frac{d\omega}{2\pi}\exp(-i\omega{}t)q(\omega)
   \quad\quad\text{and}\quad\quad
   p(t)=\int_{-\infty}^{+\infty}\frac{d\omega}{2\pi}\exp(-i\omega{}t)p(\omega)\,.
\end{eqnarray}
Integration over $\omega$ in these expressions can be carried out
with the help of Cauchy theorem. Here, we give the results for a
parabolic potential barrier,
\begin{eqnarray}\label{A5}
   q(t)=q_{0}R(t)+p_{0}S(t)+\int_{0}^{t}dt'S(t-t')\xi(t')
\end{eqnarray}
and
\begin{eqnarray}\label{A6}
   p(t)=q_{0}\left(M\Omega\right)^{2}S(t)+p_{0}Q(t)
   +\int_{0}^{t}dt'Q(t-t')\xi(t')
\end{eqnarray}
where $Q(t)$, $S(t)$ and $R(t)$ are given by,
\begin{eqnarray}\label{A7}
   Q(t)=\exp\left(-\frac{\beta}{2}t\right)
   \left(\cosh{}\overline{\Omega}t-\frac{\beta}
   {2\overline{\Omega}}\sinh{}\overline{\Omega}t\right)
\end{eqnarray}
\begin{eqnarray}\label{A8}
   S(t)=\frac{1}{M\overline{\Omega}}\exp\left(-\frac{\beta}{2}t\right)
   \sinh{}\overline{\Omega}t
\end{eqnarray}
and
\begin{eqnarray}\label{A9}
   R(t)=\exp\left(-\frac{\beta}{2}t\right)
   \left(\cosh{}\overline{\Omega}t+\frac{\beta}
   {2\overline{\Omega}}\sinh{}\overline{\Omega}t\right)
\end{eqnarray}
where $\overline{\Omega}=\sqrt{\Omega^{2}+(\beta/2)^{2}}$. The
solutions can be given in a similar manner for a parabolic
potential well. The mean values of collective variables are
obtained by taking the ensemble average of (A5) and (A6),
\begin{eqnarray}\label{A10}
   \overline{q}(t)=\overline{q}_{0}R(t)+\overline{p}_{0}S(t)
   \quad\quad\text{and}\quad\quad
   \overline{p}(t)=\overline{q}_{0}\left(M\Omega\right)^{2}S(t)+\overline{p}_{0}Q(t)\,.
\end{eqnarray}
The variances are given by,
\begin{eqnarray}\label{A11}
   \sigma_{qq}(t)=\sigma_{q0}R^{2}(t)+\sigma_{p0}S^{2}(t)
   +\int_{0}^{t}ds\int_{0}^{t}ds'S(s)S(s')2D_{0}\chi(s-s')
\end{eqnarray}
\begin{eqnarray}\label{A12}
   \sigma_{qp}(t)=\sigma_{q0}\left(M\Omega\right)^{2}R(t)S(t)
   +\sigma_{p0}S(t)Q(t)
   +\int_{0}^{t}ds\int_{0}^{t}ds'S(s)Q(s')2D_{0}\chi(s-s')
\end{eqnarray}
and
\begin{eqnarray}\label{A13}
   \sigma_{pp}(t)=\sigma_{q0}\left(M\Omega\right)^{4}S^{2}(t)
   +\sigma_{p0}Q^{2}(t)
   +\int_{0}^{t}ds\int_{0}^{t}ds'Q(s)Q(s')2D_{0}\chi(s-s')\,.
\end{eqnarray}
In these expressions, first two terms describe propagation of the
initial fluctuations of the coordinate and momentum distributions
$\sigma_{q0}$, $\sigma_{p0}$ and the last term arises from
dynamical fluctuations generated by the random force.

For calculation of the formation probability of compound nucleus,
we need an explicit expression for variance $\sigma_{qq}(t)$ of
the collective variable. In the Markovian limit, using the fact
that the correlation function behaves like a delta function,
$\chi(s-s')\rightarrow\delta(s-s')$, we obtain the known
analytical result for the dynamical part of the $\sigma_{qq}(t)$
\cite{R5},
\begin{eqnarray}\label{A14}
   \sigma_{qq}^{\chi}(t)&=&\int_{0}^{t}ds\int_{0}^{t}ds'S(s)S(s')2D_{0}\chi(s-s')
   \rightarrow\int_{0}^{t}ds\int_{0}^{t}ds'S(s)S(s')2D_{0}\delta(s-s')
   \nonumber \\
   &=&\frac{T}{M\Omega^{2}}e^{-\beta{}t}\left[
   \frac{\beta^{2}}{2\overline{\Omega}^{2}}(\sinh{}\overline{\Omega}t)^{2}
   +\frac{\beta}{2\overline{\Omega}}(\sinh{}2\overline{\Omega}t)
   -e^{+\beta{}t}+1\right]\,.
\end{eqnarray}
In the classical limit, the dynamical part of the $\sigma_{qp}(t)$
and $\sigma_{pp}(t)$ are similarly given by,
\begin{eqnarray}\label{A15}
   \sigma_{qp}^{\chi}(t)&=&\int_{0}^{t}ds\int_{0}^{t}ds'S(s)Q(s')2D_{0}\chi(s-s')
   \rightarrow\int_{0}^{t}ds\int_{0}^{t}ds'S(s)Q(s')2D_{0}\delta(s-s')
   \nonumber \\
   &=&\frac{\beta{}T}{{\bar \Omega}^{2}}e^{-\beta{}t}(\sinh{}\overline{\Omega}t)^{2}
\end{eqnarray}
and
\begin{eqnarray}\label{A16}
   \sigma_{pp}^{\chi}(t)&=&\int_{0}^{t}ds\int_{0}^{t}ds'Q(s)Q(s')2D_{0}\chi(s-s')
   \rightarrow\int_{0}^{t}ds\int_{0}^{t}ds'Q(s)Q(s')2D_{0}\delta(s-s')
   \nonumber \\
   &=&MT\frac{\beta}{\overline{\Omega}}e^{-\beta{}t}\sinh{}\overline{\Omega}t
   \left[\cosh{}\overline{\Omega}t-\frac{\beta}{2\overline{\Omega}}
   \sinh{}\overline{\Omega}t\right]+MT\left(1-e^{-\beta{}t}\right)\,.
\end{eqnarray}
For quantal calculations, introducing the Fourier transform of the
correlation function,
\begin{eqnarray}\label{A17}
   \chi(s-s')=\int_{-\infty}^{+\infty}\frac{d\omega}{2\pi}
   e^{-i\omega{}(s-s')}\tilde{\chi}(\omega)
\end{eqnarray}
we can express the dynamical part of the variance in terms of a
one-dimensional numerical integration over the frequency $\omega$
as,
\begin{eqnarray}\label{A18}
   \sigma_{qq}^{\chi}(t)=\int_{-\infty}^{+\infty}\frac{d\omega}{2\pi}
   \left|\tilde{S}_{t}(\omega)\right|^{2}\tilde{\chi}(\omega)2D_{0}
\end{eqnarray}
where $\tilde{S}_{t}(\omega)=\int_{0}^{t}dsS(s)e^{-i\omega{}s}$.
Dynamical parts of variances $\sigma_{qp}(t)$ and $\sigma_{pp}(t)$
can be evaluated in a similar manner to give,
\begin{eqnarray}\label{A19}
   \sigma_{qp}^{\chi}(t)=\int_{-\infty}^{+\infty}\frac{d\omega}{2\pi}
   \tilde{S}_{t}(\omega)\tilde{Q}_{t}^{*}(\omega)\tilde{\chi}(\omega)2D_{0}
   \quad\text{and}\quad
   \sigma_{pp}^{\chi}(t)=\int_{-\infty}^{+\infty}\frac{d\omega}{2\pi}
   \left|\tilde{Q}_{t}(\omega)\right|^{2}\tilde{\chi}(\omega)2D_{0}
\end{eqnarray}
where $\tilde{Q}_{t}(\omega)=\int_{0}^{t}dsQ(s)e^{-i\omega{}s}$.

\newpage


\newpage

\begin{figure}\hspace{-0.8cm}
\epsfig{figure=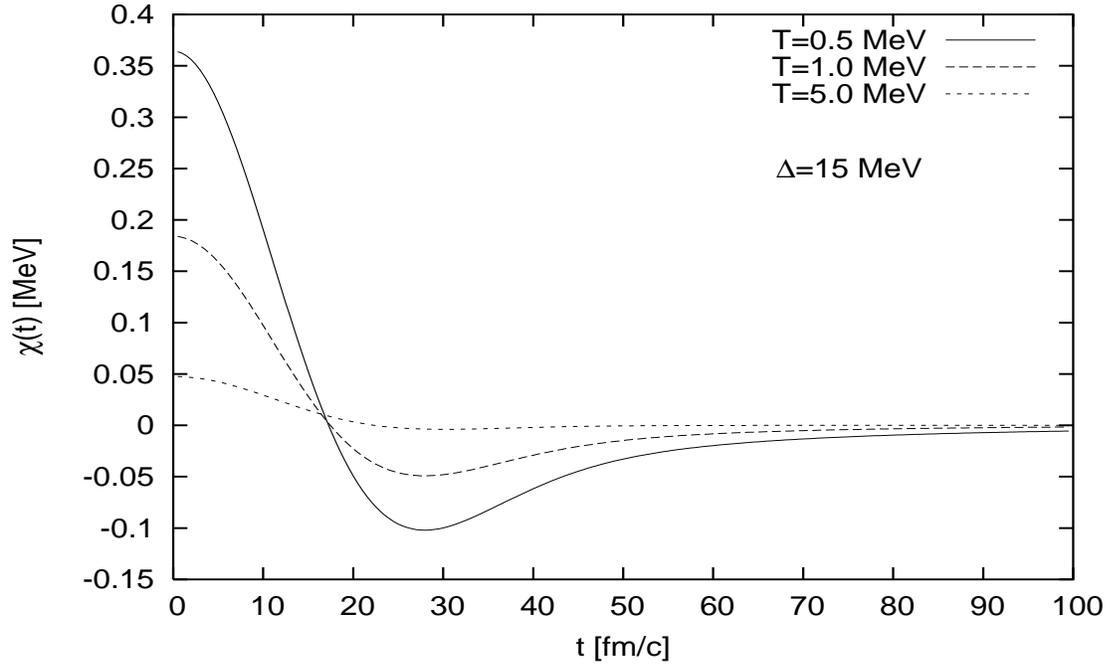,width=15cm,height=9cm} 
\vspace*{0.0cm} \caption{ The correlation function is plotted
versus time for $\Delta=15$ MeV.} \label{fig1}
\end{figure}

\begin{figure}\hspace{-0.8cm}
\epsfig{figure=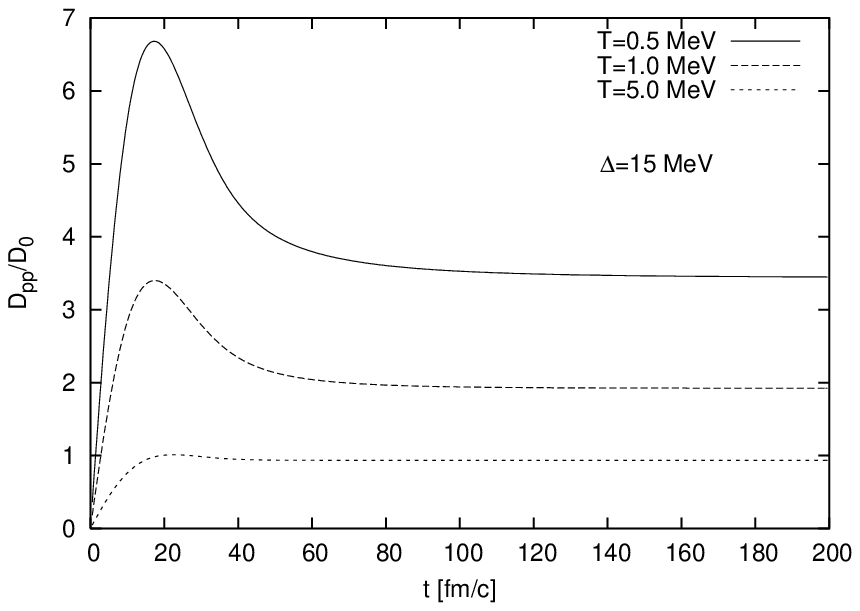,width=15cm,height=9.5cm} 
\vspace*{0.3cm} \caption{ The momentum diffusion coefficient, in
units of classical diffusion coefficient,\\ is plotted versus time
for $\Delta=15$ MeV.} \label{fig2}
\end{figure}

\begin{figure}\hspace{-0.8cm}
\epsfig{figure=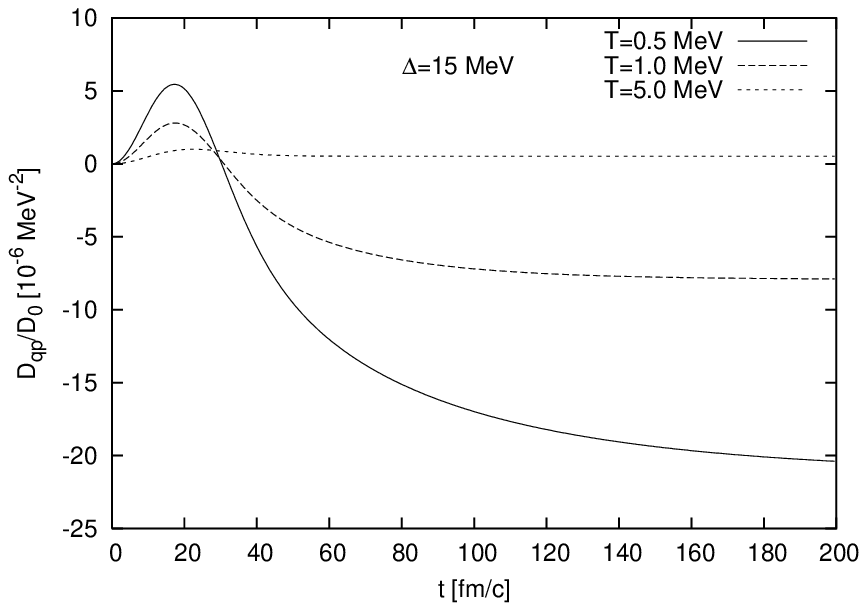,width=15cm,height=10cm} 
\vspace*{0.0cm} \caption{ The mixed diffusion coefficient, in
units of classical diffusion coefficient, is\\ \vspace*{-0.6cm}
\hspace{0.4cm} plotted versus time for $\Delta=15$ MeV.}
\label{fig3}
\end{figure}

\begin{figure}\hspace{-0.8cm}
\epsfig{figure=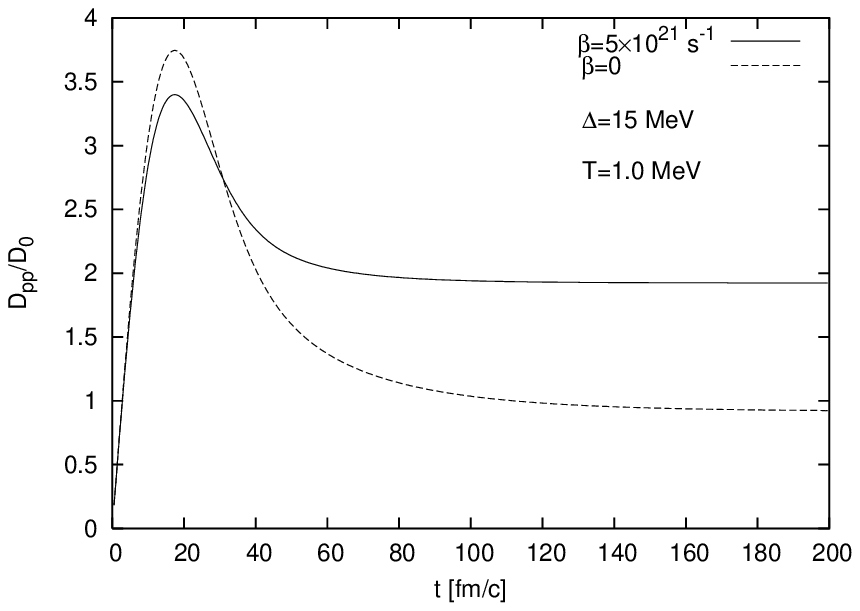,width=15cm,height=9cm} 
\vspace*{0.5cm} \caption{ The momentum diffusion coefficient, in
units of classical diffusion coefficient, is plotted versus time
for $\Delta=15$ MeV and T=1.0 MeV. The cases with and without
friction are compared.} \label{fig4}
\end{figure}

\begin{figure}\hspace{-0.8cm}
\epsfig{figure=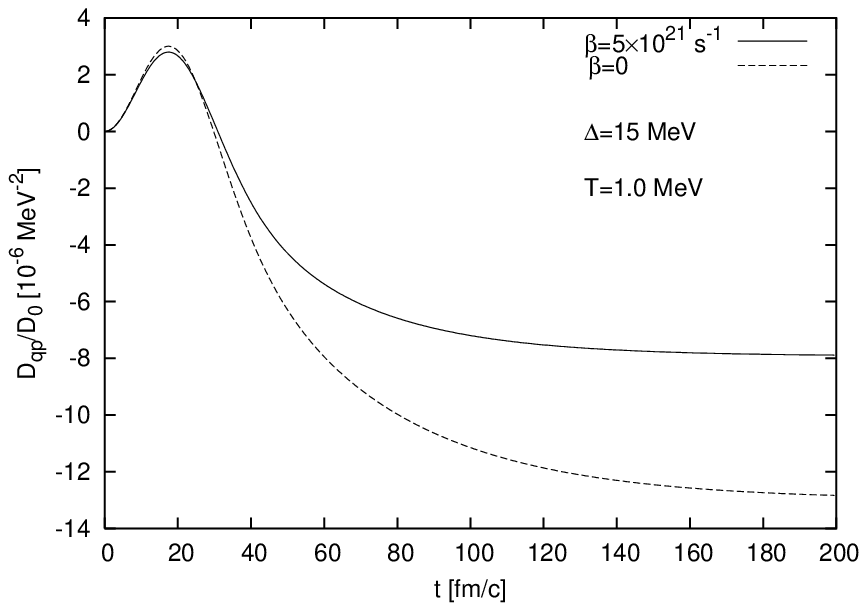,width=15cm,height=9cm} 
\vspace*{0.3cm} \caption{ The mixed diffusion coefficient, in
units of classical diffusion coefficient, is plotted versus time
for $\Delta=15$ MeV and T=1.0 MeV. The cases with and without
friction are compared.} \label{fig5}
\end{figure}

\begin{figure}\hspace{-0.9cm}
\epsfig{figure=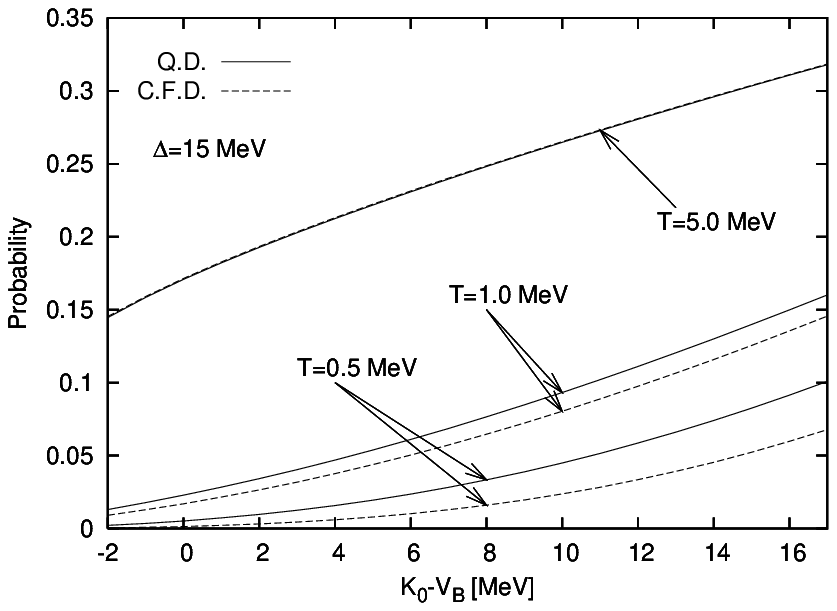,width=15cm,height=9cm} 
\vspace*{0.6cm} \caption{ The formation probability is plotted
versus initial kinetic energy minus barrier height. The result of
classical diffusion approach is compared with that of quantum
diffusion approach for $\Delta=15$ MeV.} \label{fig6}
\end{figure}


\begin{thebibliography}{99}
\bibitem{R1} P. Armbruster, Ann. Rev. Nucl. Part. Sci. 50 (2000) 411.
\bibitem{R2} Y. Arimoto, T. Wada, M. Ohta and Y. Abe, Phys. Rev. C59 (1999) 796.
\bibitem{R3} Y. Abe, Eur. Phys. J. A13 (2002) 143.
\bibitem{R4} C. Shen, G. Kosenko and Y. Abe, Phys. Rev. C66 (2002) 061602(R).
\bibitem{R5} Y. Abe, D. Boilley, B.G. Giraud and T. Wada, Phys. Rev. E61 (2000) 1125.
\bibitem{R6} S. Ayik and J. Randrup, Phys. Rev. C50 (1994) 2947.
\bibitem{R7} H. Hofmann, Phys. Rep. 284 (1997) 137.
\bibitem{R8} C. Rummel and H. Hofmann, Nucl. Phys. A727 (2003) 24.
\bibitem{R9} N. Takigawa, S. Ayik and S. Kimura, nucl-th/0203043.
\bibitem{R10} N. Takigawa, S. Ayik, K. Washiyama and S. Kimura, Phys. Rev. C69 (2004) 054605.
\bibitem{R11} C. W. Gardiner, "Quantum Noise", Springer, Berlin (1991).
\bibitem{R12} U. Weiss, "Quantum Dissipative Systems", World Scientific, Singapore (1993).
\bibitem{R13} H. Risken, "The Fokker-Planck Equation", Springer, Berlin (1984).
\bibitem{R14} P. Ring and P. Schuck, "The Nuclear Many-Body Problem", Springer, New York (1980).
\bibitem{R15} S. Ayik and C. Gregoire, Phys. Lett. B212 (1988) 269; Nucl. Phys. A513 (1990) 187.
\bibitem{R16} J. Randrup and B. Remaud, Nucl. Phys. A514 (1990) 339.
\bibitem{R17} Y. Abe, S. Ayik, P. G. Reinhard and E. Suraud, Phys. Rep. 275 (1996) 49.
\bibitem{R18} E. M. Lifshits and l. P. Pitaevskii, "Physical Kinetics", Pergamon, New
York (1981).
\bibitem{R19} V. M. Kolomietz, V. A. Plujko and S. Shlomo, Phys. Rev. C52 (1995) 2480.
\bibitem{R20} A. O. Caldeira and A. J. Leggett, Physica 121A (1983) 587-616.
\bibitem{R21} Y. Abe, C, Gregoire and H. Delagrange, J. Phys. C4 (1986) 329.
\bibitem{R22} R. L. Honeycutt, Phys. Rev. A45 (1992) 600.
\bibitem{R23} R. L. Honeycutt, Phys. Rev. A45 (1992) 604.
\bibitem{R24} Y. Abe, D. Boilley, G. Kosenko, J. D. Bao, C. W. Shen, B. Giraud and T. Wada, Prog. Theo. Phys. Sup. 146 (2002)
104.
\bibitem{R25} Y. Abe, D. Boilley, G. Kosenko and C. W. Shen,   ActaPhys. Pol. B34 (2003) 2091.

\end{thebibliography}
\end{document}